\documentstyle[epsfig,aps,prl,multicol]{revtex}
\draft
\newcommand{\br}{{\bf r}}

\newcommand{\bk}{{\bf k}}
\newcommand{\bu}{{\bf u}}
\newcommand{\cH}{{\cal H}}

\newcommand{\eps}{\varepsilon}
\newcommand{\bsigma}{{\bf {\sigma}}}
\newcommand{\grad}{{\bf {\nabla}}}

\begin{document}
\title{Dislocations and the critical endpoint of the melting line
of vortex line lattices}

\author{Jan Kierfeld and Valerii Vinokur}
\address{Argonne National Laboratory,
Materials Science Division,
9700 South Cass Avenue,
Argonne, IL 60439, USA}
\date{August 3, 1999}
\maketitle

\begin{abstract}
We develop a theory for  dislocation-mediated structural
transitions in the vortex lattice which allows for a unified description 
of  phase transitions between the three phases, the elastic vortex
glass, the amorphous vortex glass, and the vortex liquid, 
in terms of a free energy functional for the dislocation density.
The origin of  a  {\em critical endpoint} of the melting line
 at high magnetic fields, which has been recently observed
 experimentally, is explained.
\end{abstract}

\vspace{-5pt}

\begin{multicols}{2}
\textwidth42.5pc
\hsize20.7pc

Since the pioneering work \cite{N88} where the first-order flux line
lattice (FLL) melting into an entangled vortex liquid (VL) was
established, there has been a continuous development of our views of
the vortex lattice phase diagram in
high-$T_c$ superconductors \cite{blatter}.
Weak point disorder was shown to drive the vortex lattice
into a {\em vortex glass} (VG) state with zero linear
resistivity \cite{FGLV89,F89,VGexp}. Observations of hysteretic
resistivity switching and magnetization measurements
\cite{meltmag} have 
supported the first order melting of very clean lattices. A crossover
from the first order melting at low magnetic fields to a
continuous VG-VL transition has been  related to the structural
transition within the vortex solid which is described
\cite{EN96,VK98,VV98,JK98} as a topological transition between the
low-field {\em elastic} VG, free from  topological defects
\cite{KNH97,DF97} and maintaining quasi long-range translational
order \cite{Na90}, and the high field {\em amorphous} VG, where
disorder generates proliferation of dislocations \cite{KNH97}.
A simple picture of the vortex phase diagram has emerged where the
three generic phases -- VL, the high field amorphous VG,
and the low field, low temperature quasi-lattice or {\em Bragg
glass} (BrG) \cite{GL} -- are governed by the three basic energies: 
the energy of thermal fluctuations, pinning, and elastic energies.
The transition lines are determined by matching of any of the two
basic energies, and the
match of all three energies marks the tricritical point where the
first-order melting terminates \cite{VV98}.
While this simplistic picture is
supported by observations on BSCCO, it fails to describe the YBCO phase
diagram where the endpoint of the
first-order melting line appears to be separated from the point
where topological transition and melting line merge \cite{LOP}.
Furthermore,
the Lindemann-criterion approach of
\cite{EN96,VK98,VV98,JK98} can only qualitatively estimate the
location of the transitions telling nothing about the nature of the
vortex states involved.

In this Letter we undertake a step towards a quantitative theory of
the entire vortex lattice phase diagram based on a unified
description. We propose a model where all the transitions are
{\em dislocation-mediated},
and derive the free energy for an ensemble of directed dislocations
as a function of the dislocation density
in the presence of disorder and thermal fluctuations.
Each of the experimentally observed
phases is characterized by its inherent {\em dislocation
density} or, equivalently, by the characteristic dislocation spacing
$R_D$.
The elastic VG is dislocation-free and has $R_D=\infty$. The
VL can be viewed as a vortex array saturated with
dislocations such that $R_D \sim a$,
and in the amorphous VG, $R_D \sim R_a$, where $R_a$ is
the so-called positional correlation length on which
typical vortex displacements are of the order of  the lattice spacing 
 $a$ \cite{blatter}. 
Each phase corresponds to one of the {\em local} minima 
in the dislocation ensemble free energy, and  dislocation densities
in these minima  represent the {\em equilibrium} 
dislocation densities in the corresponding phases. 
The global minimum corresponds to the thermodynamically stable phase 
under the given conditions, phase 
transitions occur when  
 two local minima exchange their role as global minimum.
This mechanism for the transitions enables us to 
{\em derive} Lindemann-criteria both for the locations of
the thermal melting line and the disorder-induced instability line
of the BrG.
The characteristic scale set by the
mean distance between free dislocations offers a
natural explanation of the critical endpoint of the first-order melting line.
While at low magnetic fields $R_a \gg a$ and the amorphous VG
appears to contain significantly less dislocations than the VL, 
at higher field where $R_a = a$
the two phases become thermodynamically
{\em equivalent} and the
first order melting line terminates.

A free energy for the dislocation degrees of freedom governing
phase transitions is derived from the vortex lattice elasticity theory.
Dislocations in the FLL can be both of  screw or edge type, 
but in either case they  are confined to the gliding
plane spanned by their Burger's vector ${\bf b}$ and the magnetic field
\cite{NM90}.
The single dislocation energy consists of the core energy $E_c$ and of the
logarithmically diverging contribution from the long-range elastic
strains \cite{hirth}.
Accordingly, the dislocation ensemble is modeled
as an array of elastic lines with a
long-range Coulomb-like interaction.
The energy penalty for Burger's vectors with $b>a$ leads to an effective 
hard-core repulsion of dislocations which conspires with the planarity 
constraint to limit the entropy gain of dislocations. 
This favors  {\em directed} dislocations threading the entire sample
\cite{kardar} rather than large 
dislocation {\em loops} \cite{loops}.
A single directed dislocation line is parameterized by its 
displacement field $u_D(z)$ and described by the Hamiltonian
\begin{equation}
\cH_{D}[u_D] = \int dz \left( E_{s} + \frac{1}{2} \epsilon_D (\partial_z
u_D)^2 \right)
\label{elasticD}
\end{equation}
where the stiffness
$\epsilon_D \propto \ln{\left(1/k_za\right)}$
has a logarithmic dispersion from the long-range interaction and $E_s$
is the self-energy of a straight dislocation.
For thermal melting,
the short wavelength limit, $k_z \sim 1/a$, is relevant.
After rescaling
$z =\tilde{z} \frac{1}{2}\sqrt{c_{44}/c_{66}}$
dislocation energies become {\em isotropic} with
$\epsilon_D = E_c = c_{D} K b^2/4\pi$ ($c_D \approx 1$)
and
$E_{s} = E_{c} + K b^2/4\pi \ln{\left(L/a\right)}$
where $K=\sqrt{c_{44}c_{66}}$ is the {\em isotropized elastic
constant} ($c_{44}$ and $c_{66}$ are the tilt and shear moduli of the
vortex lattice, respectively. 
$L$ is the system size in the transverse direction).
Note that the Peierls barrier $W_p$
and the associated ``kinking'' \cite{hirth}
of dislocation lines can be neglected near the melting transition.
It can be shown that kinks are
irrelevant above the temperature 
$T_k \sim a \sqrt{W_d \epsilon_d}$. Since we find 
 $W_p \lesssim 10^{-4} E_c$, $T_k$ is much {\em lower} than $T_m$.
Therefore,
the basic length scale along the 
magnetic field is solely set by the competition of
FL tilt and shear and given by
$a_z \approx \sqrt{c_{44}/c_{66}}/2$
($a_z \approx a$ in the rescaled system).
The free energy of a single dislocation can be readily calculated from the
partition sum $Z_D = \int{\cal D}u_D \exp{(- \beta\cH_{D}[u_D])}$ by
Gaussian functional integration and consists of
the core energy, the long-range strains elastic energy,
and the entropy term:
\begin{eqnarray}
\frac{F_D(L)}{L_z} &=& E_c + \frac{Kb^2}{4\pi}
\ln{\left(\frac{L}{a}\right)}
-T \frac{1}{2a_z}
\ln{\left(1+ \frac{2\pi T a_z}{\epsilon_D a^2}\right)}
\label{singleFD}
\end{eqnarray}
A second order
phase transition due to the formation of a {\em single}
dislocation is prohibited by the logarithmic divergence of 
its elastic energy. 
However, in an {\em ensemble} of dislocations, this divergence
is screened out on the distance $1/2\rho a$, where  $2\rho$ is the
dislocation 
density (the ensemble is topologically neutral
to avoid the accumulation of stresses).
An additional entropy cost ($\propto \rho^3$)
comes from the steric repulsion, and
the resulting free energy density then reads
\begin{eqnarray}
f(\rho) &=& 2\rho \left( E_c -T \frac{1}{2a_z}
\ln{\left(1+ \frac{2\pi T a_z}{\epsilon_D a^2}\right)} \right)
+ \nonumber\\
&& ~+ 2\rho \frac{Kb^2}{4\pi}
\ln{\left(\frac{1}{2a^2\rho}\right)}
+ \rho^3 \frac{\pi^2}{3} \frac{T^2a^2}{\epsilon_D}.
\label{f}
\end{eqnarray}
$f(\rho)$ can be derived in a more rigorous manner by mapping
dislocations 
onto a quantum system of 2D Fermions with Coulomb
interaction \cite{YI88,tobe}.
At high dislocation densities, the screening mechanism in  (\ref{f}) 
is purely entropic in  nature and equivalent to the 
formation of an exchange-hole in the 2D fermionic system.
A {\em first order melting}
 following from  (\ref{f}) occurs at 
$ T_m \approx 1.5 E_c a_z \approx 0.15 Ka^3$,
which is equivalent to melting according  to the Lindemann-criterion
with a Lindemann-number $c_L\approx 0.2$; in 
good agreement with experimental and numerical results.
At the melting transition dislocations proliferate with a
{\em high} density
$\rho_m \approx 0.3 a^{-2}$, hence
the VL  is saturated with dislocations.
The presented scenario does not require
a {\em simultaneous}  proliferation of disclinations, however, the latter 
are likely to appear at the high dislocation densities involved.

In the presence of a random potential $V_{pin}(\br)$,
the collectively pinned {\em dislocation-free} vortex array
passes through three different scaling regimes:
(i)
Small scales where vortex displacements $u$ are 
smaller than the coherence length $\xi$ 
 and perturbation theory applies \cite{LO}.
(ii)
Intermediate scales where $\xi \lesssim u \lesssim a$ and
disorder potentials seen by different FLs are effectively {\em
uncorrelated}.  This regime is
captured in so-called {\em random manifold} (RM) models \cite{blatter,GL},
leading to a roughness
$\tilde{G}(\br)=\overline{\langle (\bu(\br)-\bu(0))^2 \rangle}
\approx 4 (a/2\pi)^2 (r/R_a)^{2\zeta_{RM}}$
where $\zeta_{RM} \approx 1/5$ for the $d=3$
dimensional RM with two displacement components.
The crossover scale to the asymptotic behaviour is the
{\em positional correlation length} $R_a$ where
the average displacement is of the order of the FL spacing:
$u \approx a/2\pi$ \cite{note}.
(iii)
The asymptotic {\em Bragg glass}
regime where the $a$-periodicity of the FL array becomes important
for the coupling to the disorder and the array is effectively
subject to a {\em periodic} pinning potential with period $a$ \cite{Na90}.
Here the {\em logarithmic} roughness
$\tilde{G}(\br) \approx 4 (a/2\pi)^2 \ln{\left(e r/R_a\right)}$,
i.e., $\zeta_{BrG} = {\cal O}(log)$ \cite{Na90,GL}  takes over.

In a disordered system at $T=0$ the mechanism for dislocation
proliferation is fundamentally different from the thermal melting
discussed before.
While thermal melting is driven by the {\em entropy gain} from unbinding
dislocations, at $T=0$ the  FLL benefits energetically 
 from
adjusting itself to the disorder, and dislocation
proliferation leads to {\em disorder energy gain}
through the dislocation degrees of freedom.
 It has been shown in Refs.~\cite{KNH97,DF97}
that the BrG phase in 3D
is {\em stable} with respect to dislocation formation.
Instabilities  arise, however, from
the sub-asymptotic regimes [in Ref.~\cite{KNH97}
this has been partly taken into account by considering
displacements in the dislocation core].
To handle analytical difficulties and to provide a unified treatment
through all scaling regimes, we develop an  approach to the 3D problem 
based on
an {\em effective random stress model} which has the same
displacement correlations as the full non-linear disordered
model but allows for a {\em separation} of dislocation and elastic
degrees of freedom.
This  idea is motivated by the renormalization group (RG)
for the 2D BrG which explicitly shows it
renormalizes asymptotically into a random stress model \cite{CO82} and 
has been used in Ref.~ \cite{ZLF99} to show the
{\em instability} of the 2D BrG with respect to dislocations.
For simplicity
we consider a {\em uniaxial} model (in the incompressible
limit $c_{11}\gg c_{66}$) which yields the same
dislocation energetics as the isotropized two-component model.
The Hamiltonian is
\begin{equation}
\cH[\bu] = \int_\br \left\{ \frac{1}{2} K (\grad u)^2 +
\bsigma \cdot \grad u \right\}
\label{Hranstress}
\end{equation}
where
$\bsigma(\br)$ is the {\em random stress field }
which we assume to be Gaussian distributed with
a second moment $\overline{\sigma_i(\bk)\sigma_j(\bk')}=
\delta_{ij} \Sigma(k) (2\pi)^3 \delta(\bk+\bk')$ characterized by the
function $\Sigma(k)$ in Fourier space.
The effective random stresses 
causing displacements with the same
(2-point) correlations as those for the
RM or BrG regime are
\begin{eqnarray}
\Sigma(k) &=& \left\{
\begin{array}{ll}
\mbox{BrG:}
& \frac{1}{2} K^2k^{-1}a^2 \\
\mbox{RM:}
& B_{1/5} K^2k^{-1} a^2(k R_a)^{-2/5}
\end{array} \right.
\label{GRMBG}
\end{eqnarray}
with a numerical constant $B_{1/5}$ depending only on
the roughness $\zeta_{RM}=1/5$ (the exact
crossover between the two regimes is non-trivial \cite{GL}).
The validity of the random stress model is
well-established in 2D.
Besides, the functional RG treatment of the BrG in $d=4-\epsilon$
dimensions shows
that displacements asymptotically obey  Gaussian statistics
up to the first order in $\epsilon$ \cite{emig},
which can always be modeled by an effective random stress field.
Similarly, in a real-space RG analysis \cite{VF84},
the relevant fixed point is perturbative in $\epsilon$ and an
analogous argument applies.

Starting from (\ref{Hranstress})  the
energy for an ensemble of curved dislocation lines
with a Burger's vector density
${\bf b}(\br)$ 
in the disordered system can be calculated
(for example along the lines of Ref.~\cite{LNRS96}).
In the random stress model the Hamiltonian
{\em decouples} into the elastic part and a dislocation part:
\begin{eqnarray}
{\cal H}_D[{\bf b}] &=& \int_\br \int_{\br'} \frac{K}{2}
{\bf b}_\br \cdot {\bf b}_{\br'} G_0(\br-\br')
+ \int_\br {\bf b}_{\br} \cdot{\bf g}_{\br}
\label{HDrandom}
\end{eqnarray}
where $G_0(\br) = 1/(4\pi r)$ is the 3D Green's function.
Here ${\bf g}(\br)$ is an effective
{\em random potential} for dislocations lines
defined by the transversal part of $\bsigma$ through
$\grad \times {\bf g} = \bsigma_T$ (cf.~\cite{ZLF99}).
This energy contains the long-range elastic energy $E_{s}$
of dislocations in
the first term and in the stochastic second term the disorder energy
$E_{dis}$ of the dislocation which allows dislocations to gain
energy by optimizing their paths. The disorder energy is completely
determined by the FL
displacement correlations through
$ \overline{g_i(\bk)g_j(\bk')} = \delta_{ij} K \Sigma(k)k^{-2} (2\pi)^3
\delta(\bk+\bk')$
in the different regimes given by (\ref{GRMBG}).
For a single directed dislocation line,
the Hamiltonian (\ref{HDrandom}) reduces to the problem of a directed
elastic line (with a logarithmic dispersion, see above)
in a random potential that is long-range correlated due to
(\ref{GRMBG}).
For a directed dislocation of length $L_z$ and
confined in the transversal direction to a scale $L$,
 the elastic energy is again given
by (\ref{singleFD}) with $T=0$.
The mean square disorder energy fluctuations are
\begin{eqnarray}
\overline{E_{dis}^2(L_z,L)} &=& \left\{
\begin{array}{ll}
\mbox{BrG:} & c_{BrG}E_D^2 L_z L \\
\mbox{RM:} & c_{RM} E_D^2 L_z L
\left(\frac{L}{R_a}\right)^{2/5}
\end{array} \right.
\label{Edis2}
\end{eqnarray}
with numbers $c_{BrG}, c_{RM} = {\cal O}(1)$.
These expressions give an estimate of the {\em typical} disorder energy a
dislocation line can gain. They neglect
{\em rare fluctuations} in the energy gain from optimally positioning
the dislocation in the transversal plane which
give logarithmic corrections $\sim{\cal O}(\ln{L})$ \cite{DF97}.
The optimal path of the dislocation will be
{\em rough} $u_D \sim L_z^{\zeta_D}$ with an exponent $\zeta_D$.
The roughness can be obtained by
a Flory argument that equates the elastic energy from the deformation
$\epsilon_D(L) u_D^2/L$ (with a logarithmically
dispersive $\epsilon_D \sim \ln{L}$ on large
scales) and the disorder energy
$\overline{E_{dis}^2(L_z=L,L)}^{1/2}$ on {\em one} large length scale set
by the fluctuation wavelength $L$. This yields
$\zeta_{D}(BrG) = 1 - {\cal O}(log^{1/2})$ and
$\zeta_{D}(RM) = \frac{11}{10} - {\cal O}(log^{1/2})$
where logarithmic corrections come from the
dispersion of the stiffness and rare fluctuations.
Since  $\zeta_{D}(BrG)\le 1$, the BrG
appears to be marginally {\em stable} against penetration of a {\em
single} directed dislocation whereas $\zeta_{D}(RM)>1$ such that
the random manifold is clearly {\em unstable}. 
Note that the scaling arguments of Ref.~\cite{DF97}
taking into account rare fluctuations
give the same result as the simple
Flory argument neglecting rare fluctuations.
We argue that the sub-asymptotic instability of the FLL in the RM
regime on
scales $L< R_a$, combined with the asymptotic stability in the BrG
regime for scales $L>R_a$, leads to a  disorder-induced dislocation
proliferation via a {\em weak first order} phase transition.
The characteristic dislocation density $\rho_c \sim R_a^{-2}$ at
the transition is given just by the crossover scale $R_a$.
The discontinuities in this transition are small
and may eventually disappear
if  the length scale $R_a$ becomes of the
order of typical sample dimensions.
Note also that, qualitatively,
this result is based {\em only} on the fact that the instability sets
in within a sub-asymptotic regime; the random stress model is only
used to quantify our findings.
It enables us to estimate typical
free energy {\em minima} for ensembles of
dislocation lines with rough
optimized paths at $T=0$.
The screened long-range elastic energy density for
a (neutral) dislocation ensemble with density $2\rho$ is given by
$e_{D}(\rho) = 2\rho \left( E_D + (Kb^2/4\pi)
\ln{\left(1/\sqrt{2}a\rho^{1/2}\right)} \right)$ as in (\ref{f}) at $T=0$.
Dislocations are confined to a transversal scale $R_D\simeq
\rho^{-1/2}$ set by the distance to the next dislocation,
hence they optimize their disorder energy gain
on each longitudinal scale $L_z \simeq R_D$ {\em independently}.
Using (\ref{Edis2}) with $L_z=L=\rho^{-1/2}$ for the BrG regime
($\rho< R_a^{-2}$) and the RM regime
($\rho> R_a^{-2}$), we can estimate
the corresponding minimal free energy densities (not considering
logarithmic corrections from rare fluctuations)
\begin{eqnarray}
f(\rho) &\approx& e_{D}(\rho)
- \left\{
\begin{array}{ll}
\mbox{BrG:} & 2A_{BrG}  E_c \rho
\\
\mbox{RM:} & 2A_{RM}\frac{E_c}{a^2} (\rho a^2)^{\frac{9}{10}}
\left(\frac{a}{R_a} \right)^{\frac{1}{5}}
\end{array} \right.
\label{fdis}
\end{eqnarray}
with numbers $A_{BrG}, A_{RM} = {\cal O}(1)$.
When both results in (\ref{fdis}) are combined
one indeed finds a
local minimum in the free energy density
at $\rho \approx R_a^{-2}$
that characterizes an amorphous VG phase.
Over a wide range of magnetic fields the dislocation density in
the amorphous VG is much {\em lower} than in the
VL for which we have found $\rho \approx 0.3 a^{-2}$ above.
The elastic BrG phase looses stability with respect to
dislocation proliferation and a transition
into an amorphous VG phase
if the local minimum at $\rho \approx R_a^{-2}$
becomes the global free energy minimum.
This occurs via a
weak first order transition when
$R_a/a = C$ with a ``Lindemann-number'' $C= {\cal O}(10)$.
This is {\em identical} to the Lindemann criterium obtained in
Refs.~\cite{KNH97,JK98} including 
the numerical value of $C$ \cite{note}. As it was shown in
Ref.~\cite{JK98} it is equivalent to a Lindemann-criterion
$\overline{\langle (u(a)-u(0))^2 \rangle} = c_L^2 a^2$ in its more
familiar form (see \cite{EN96}) with $c_L = \frac{1}{2\pi} C^{-1/5}
\approx 0.1$.

\begin{figure}[th]
\epsfxsize=3.2in
\epsffile{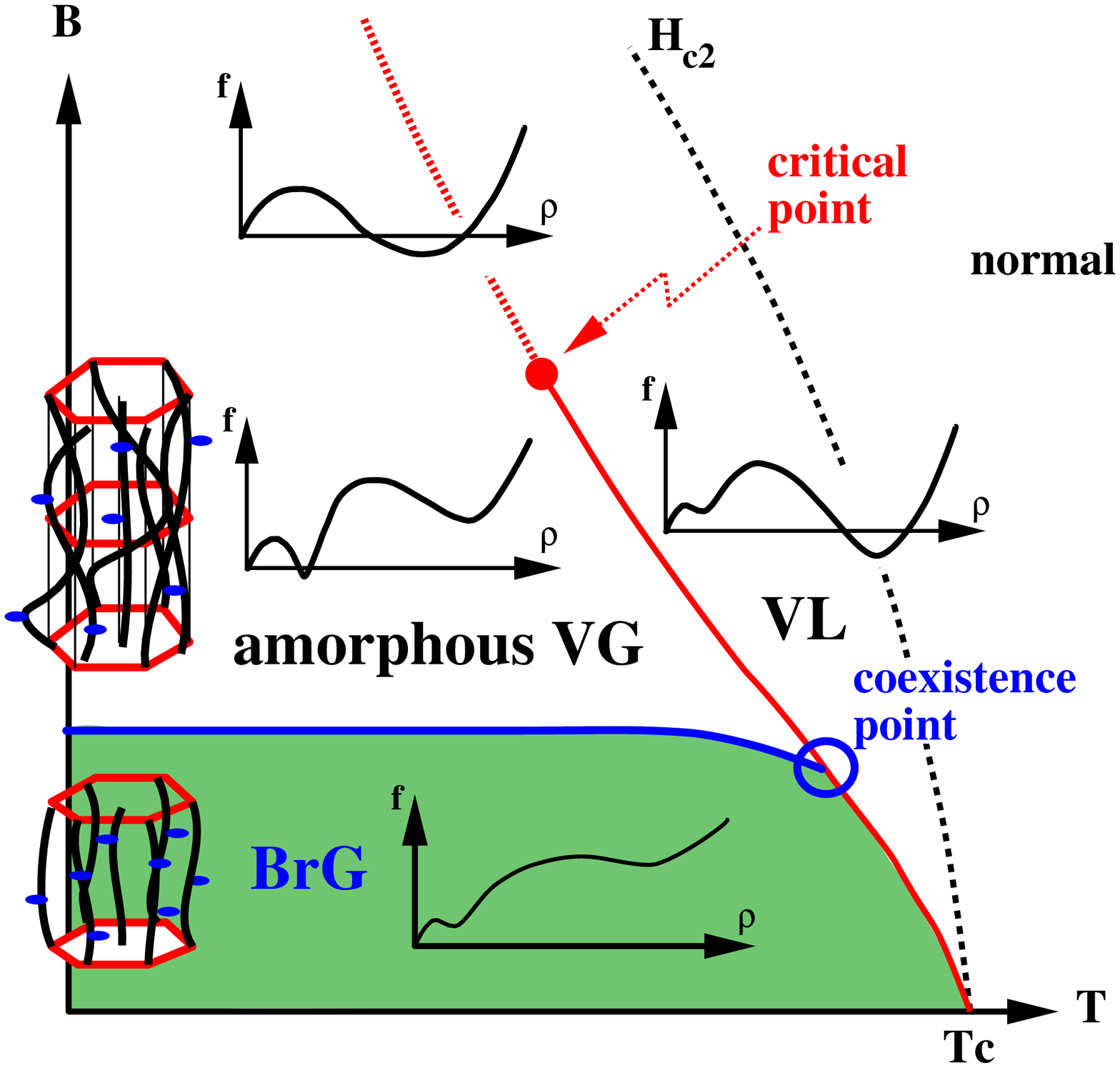}
{FIG.~1.~Schematic phase diagram of
YBCO. Insets show typical free energy densities $f$
of a dislocation ensemble as function of the dislocation density $\rho$.}
\end{figure}

We have identified {\em three} possible minima for the free energies 
(\ref{f}) and (\ref{fdis}):
(i) The dislocation-free
minimum at $\rho=0$ which is stable in the elastic BrG phase
at low $T$ and low $H$. (ii) The minimum at $\rho \sim a^{-2}$
that becomes stable in the disorder-free case for high $T$ in the
VL.
(iii) A minimum at $\rho \sim R_a^{-2}$ which is realized
in the amorphous VG.
Combining our results for the thermal melting and the
disorder-induced ``melting'', we have obtained
a qualitative theory for the {\em entire} phase diagram of the vortex
matter.
The experimentally observed {\em critical endpoint} 
of the first order melting line finds a
natural explanation.
At elevated fields the positional
correlation length $R_a$ {\em decreases} \cite{JK98}
and finally reaches $R_a \sim a$
such that the two free energy minima of the VL and the
amorphous VG {\em merge}. Both these phases
become thermodynamically indistinguishable and have identical
{\em equilibrium} lattice order. Above the critical endpoint
there might still exist a {\em dynamic} transition (or crossover)
which involves the thermal depinning of dislocations,
similar to the well-known thermal depinning transition of, for example, a
single pinned vortex line.
The critical endpoint is located at the magnetic induction
determined by the condition $R_a \approx 2 a$ (which is again equivalent
to a Lindemann criterion but with $C_{cp} \approx 2$ or $c_L \approx
0.14$). This gives the estimate
$B_{cp}/H_{c2} \approx (2\pi)^{-13/3}
\left(\delta/\eps\right)^{-10/9}
C_{cp}^{-16/15} \sim {\cal O}(10^{-1})
$
where we followed Ref.~\cite{blatter} and
introduced a dimensionless disorder strength $\delta/\eps
\sim 10^{-3}$ for high-$T_c$ materials (cf.~\cite{JK98}).
This value is approximately a factor 10 higher than the
instability field $B_{BrG}$ of the BrG and thus
the ``coexistence point'' where the 
topological transition line
ends in the first order melting line and all three phases
-- elastic BrG, amorphous VG, and VL --
can coexist.

We thank A.E.\ Koshelev, H.\ Nordborg, and D.\ Lopez for stimulating
discussions,  M.\ Kardar for bringing the article of T.~Bohr to our
attention, and D.\ Blair for a critical reading of the manuscript.
 This work was supported by Argonne National Laboratory
through the U.S.\ Department of Energy, BES Materials Sciences,
under Contract No.\ W-31-109-ENG-38, and by NSF-STC under Contract 
No.\ DMR91-20000 STcS.
J.\ K.\ acknowledges support from the Deutsche Forschungsgemeinschaft
under Grant No.~KI~662/1.

\end{multicols}

\end{document}